\documentclass{elsart}
\usepackage{natbib}

\usepackage{graphicx}
\usepackage{epsfig}
\usepackage{caption2}
\usepackage{amsmath}
\usepackage{amssymb}

\newcommand{\eps}{\varepsilon}
\newcommand{\w}{\omega}

\newcommand{\prb}[1]{{\em Phys. Rev. B}, {\bf{#1}}}
\newcommand{\prl}[1]{{\em Phys. Rev. Lett.}, {\bf{#1}}}

\begin{document}

\begin{frontmatter}
\title{Surface plasmon resonances in the absorption spectra of nanocrystalline cupric oxide.}

\author[imp]{B.A. Gizhevskii}
\author[imp]{Yu.P. Sukhorukov}
\author[imp]{N.N. Loshkareva}
\author[usu]{A.S. Moskvin}
\author[usu]{E.V. Zenkov}
\author[snezh]{E.A. Kozlov}

\address[imp]{Institute of Metal Physics, Ural Division of Russian Academy of Sciences,
620219 Ekaterinburg, Russia}
\address[usu]{Ural State University, 620083 Ekaterinburg, Russia}
\address[snezh]{Russian Federal Nuclear Center -
              Institute of Technical Physics, 456770, Snezhinsk, Russia}

\begin{abstract}
Optical absorption spectra of nanocrystalline cupric oxide CuO
samples, obtained using the converging spherical shock wave
procedure, reveal significant spectral weight red-shift as compared
with spectra of single crystalline CuO samples. In addition, some of
these samples manifest the remarkable temperature-dependent
"peak-dip-hump" feature near 1.3-1.6 eV. The minimal model suggested
to explain both effects implies the nanoceramic CuO to be a system
of two species of identical metallic-like droplets with volume
fractions $p_1 \gg p_2$ and damping parameters $\gamma _1 \gg \gamma
_2$, respectively, dispersed in an effective insulating matrix. In
other words, both effects are assigned to the surface plasmon (Mie)
resonances due to a small volume fraction of metallic-like nanoscale
droplets with Drude optical response embedded in the bare insulating
medium.
 Simple effective medium theory is shown to provide the reasonable
description of the experimental spectra.
\end{abstract}


\end{frontmatter}

%\pacs{78.40.Pg, 78.66.Nk, 78.66.Vs}
% 78.40.Pg   Disordered solids
% 78.66.-w   Optical properties of specific thin films
%      78.66.Nk   Insulators
%      78.66.Vs   Fine-particle systems

\section{Introduction}

In the past decade a good deal of interest has renewed to the cupric
oxide CuO, generally recognized as  the prototype material of a
broad family of strongly correlated (SC) oxides. It is now believed,
that a considerable amount of data, obtained by means of different
experimental techniques gives evidence of an intrinsic electron
inhomogeneity of these materials. The stripe scenario gained the
prime importance in the context, and the observation of stripes in a
representative set of HTSC's and non-superconducting oxides,
including CuO \cite{prl85} was reported.

The tendency toward the phase separation and insulator-to-metal
transition upon the chemical substitution  is observed in many
cuprates. In optical experiments, different stages of this process
manifest itself in a marked increase of the infrared (IR) absorption
within the dielectric gap, and the subsequent formation of the
low-energy tail due to the itinerant carriers. Such an evolution is
clearly seen in doped HTSC's and manganites and a number of other SC
oxides \cite{sces93}. In a series of papers \cite{physB} we proposed
an unified concept of the inhomogeneity in doped SC oxides, treating
it as a system of polar electron and/or hole centers, the preformed
nanoscale hole-rich and hole-poor regions, where the breakdown of
the uniform charge distribution and the stable disproportionation is
favoured by the Jahn-Teller nature of Cu (Mn) ion. The quantitative
description of optical absorption spectra of basic SC oxides such as
La$_{2-x}$Sr$_x$CuO$_4$ \cite{lumin}, La$_{1-x}$(Ca,Sr)$_x$MnO$_3$
\cite{fttMn,jetp03} was made possible within the approach and the
complex structure of IR spectra of these systems have been
interpreted as the manifestation of geometric (surface plasmon)
resonances. Moreover, it seems probable that SC oxides which ground
state is a result of competition between two phases have to contain
a certain volume fraction of the droplets of competing phase.

While the nature and the existence itself of this {\it intrinsic}
inhomogeneity in SC oxides remains a hot debated issue, there are a
number of experimental technique to induce the electronic
inhomogeneity extrinsically. The comparison of the physical
properties of SC oxides with a controllable  {\it extrinsic}
inhomogeneity and that of doped SC oxides is of especial interest,
since it enables to draw the conclusions about the (presumably)
intrinsic inhomogeneity in the latter and makes it possible to get
insight into the underlying physics of the oxides and the
insulator-to-metal transition. The similarity of the effects of
intrinsic and extrinsic inhomogeneity, observed under different
conditions, could support the existence of the competing phases,
that  always present in the SC oxides and evolving, when the doping
and/or external influence suppress the stability of the parent
ground state.

Below we address  simple monoxide CuO which is believed to be a
relevant model system for a wide series of SC cuprates.
 Earlier on we have performed a detailed study of
 optical absorption of nominally pure stoichiometric  single crystalline CuO
  samples, as well the samples, exposed to the fluence of the neutrons,
   electrons, and He$^+$-particles \cite{CuN,CuEl,CuA}. Here, as in the case of nonisovalent
chemical substitution in other cuprates, the accumulation of the
defects leads to the redshift of the fundamental band edge and the
appearance of new absorption spectral features within the dielectric
gap. Hereafter we present the results of the optical studies of a
series of high density polydisperse CuO nanoceramics. This material
belongs to the nanocrystal oxides, which are the metastable highly
imperfect systems with generally substantial stoichiometry
violation, determined by the sample preparation procedure. They are
likely to exhibit a strong spatial nonuniformity with a broad
distribution of concentrations and the physical properties of the
inhomogeneities, that makes it possible to speak about the extended
phase separation. The optical data relevant to the systems under
consideration remain scanty because of general difficulties of the
investigations of strongly disordered materials, and on the other
hand, due to the problems in the preparation of dense granular
samples required for optical experiments. Thus, the optical studies
of the nanocrystalline cupric oxide is of especial interest as a
probe of the peculiarities of its energy spectrum and as a promising
method of the detection of the phase inhomogeneity. The rest of the
paper is organized as follow. In the next section the experimental
details and obtained results are presented. The theoretical
interpretation and conclusions are given in Secs. III. and IV.

\section{Experiment}

The nanoceramic CuO specimens have been obtained by the processing
of a standard polycrystalline samples with the mean size of the
grain varying from 5 to 20 $\mu$m \cite{FMM} using the converging
spherical shock wave procedure \cite{kozlov}. The noticeable
strength and high density of the ceramics (that amounts to 99 $\%$
of that of the theoretical close-packed structure) makes it
appropriate for the preparation of thin platelets of 40 - 70 $\mu$m
in thickness, suitable for the measurements of optical absorption.
Three samples, henceforth denoted as No 1, 2 and 3, with the mean
size of the grain equal to 20, 60 and 100 nm, respectively, have
been investigated, the granularity being estimated using the
scanning tunnel microscope and by the broadening of X-ray
diffraction lines (XRD). The XRD data and the analysis of O
K$\alpha$ X-ray emission spectra (XES) attested all samples to be
monophase \cite{FTT}. The annihilation radiation angular
correlations (ACAR) analysis  gave evidence of a considerable
concentration of the oxygen vacancies and their agglomerates at the
boundaries of crystalline grains \cite{11a}.
%These defect
%aggregates can catalyse the nucleation of the novel phases with an
% enhanced low-energy optical %response.

The measured absorption spectra $K(\w)$ are shown in Fig.
\ref{fig1}, where the data are normalized to $K$ (2.0 eV). It can be
seen, that the spectra of nanogranular samples differ in many
respects from those of CuO single crystals \cite{CuEl}.  The main
peculiarity of the present spectra is a dramatic transfer of
spectral weight from the fundamental band to lower energies. Similar
effect has been previously observed also in the spectra of CuO after
the electron and ion irradiation \cite{CuN,CuEl,CuA}. More
pronounced spectral redistribution in our case can be regarded as an
evidence of a higher degree of the structure imperfection and
inhomogeneity of the nanoceramics. While the fundamental band edge
in pure CuO is known to be 1.45 eV for room temperature \cite{JETP},
the corresponding quantity for the present inhomogeneous
nanoceramics can hardly be ascertained.

The most striking peculiarity of the spectra (Fig. \ref{fig1}) is a fine oscillating
structure, smooth or quite pronounced, that emerges near the fundamental band edge.
This is most clearly observed in the spectrum of the sample  2, where a remarkable
''peak-dip-hump'' feature with three peaks resolved at 1.33, 1.55 and 1.82 eV.
 This
feature manifests a puzzling temperature dependence, most distinctly observed
 in the
spectrum of the sample 2 (panel B of Fig. \ref{fig1}): while clearly
 resolved
 at $T=$
295 K, it completely disappears upon cooling to 80 K. It is worth
noting that this behaviour is
 obviously incompatible with the excitonic picture.

While the shape of the spectral profile (Fig. 1) is ascertained with
a high precision, the absolute value of the absorption can hardly be
displayed as reliably. E.g., the intensity of the pronounced peak at
1.33 eV is of about 2100 cm$^{-1}$. Such a small absorption near the
fundamental band edge is at variance to the values derived  from the
typical complex dielectric permittivity of insulating cuprates like
CuO, La$_2$CuO$_4$, from where the values of $K \,\sim\,$ 10$^5$
cm$^{-1}$ can be derived \cite{uchida}. However, the magnitude of
the early reported absorption coefficient of the CuO polycrystals
\cite{marabelli} is in a reasonable agreement with the present data.
We believe, that such a discrepancy is primarily related to the
differences of the techniques based on the transmittance and
reflectance  analysis. While inessential for an ideal uniform
medium, it may become crucial and unavoidable for strongly
disordered systems, where the surface and bulk properties of the
sample can differ due to the variation of the concentration and the
structure of the inhomogeneity. In fact, it is only possible to
speak about the effective optical characteristics of inhomogeneous
system, where the ''physical'' contribution, related to the energy
band structure, can hardly be separated from the ''geometrical''
one, that reflects the subtleties of the the nanoscale structure of
a particular sample. Thus, for example, it is difficult to derive
the dielectric permittivity of strongly inhomogeneous film from the
measurements of its optical density, because the effective optical
path length can be strongly elongated due to multiple reflections
accompanied by an intricate frequency dependence.

\section{Discussion}

Proceeding to the theoretic analysis of the findings of Sec. II, we firstly address
the strong red-shift effect in the spectra of CuO nanoceramics,  that could  hardly be explained by simple manifestation of
the lattice imperfection and points to a strong rearrangement of electronic structure.
 Taking
into account the nanoscopic texture of the samples we can speculate
that the anomalous spectral weight shift effect evidences the
appearance of the highly polarizable, or metallic-like nanoscopic
droplets with a pronounced low-energy optical response.  The
question of their microscopic nature is involved and is out of the
scope of the present paper. Below we shall approximate their optical
response by conventional Drude model with a  rather wide
distribution of effective plasma frequencies $\omega _p$  and
relaxation rates $\gamma$. Thus we assume   the nanoceramic CuO
samples to form a disordered
 insulating matrix with  highly polarizable inclusions having a Drude-like optical
 response.  We shall term them as
metallic-like droplets, although, properly speaking, the character of their optical
response near zero frequency and the dc conduction properties cannot be fixed proceeding from the available experimental data.

The optical properties of such a metal-insulator composite can be described in frames
of the effective medium theory \cite{emt}. The theory predicts the unconventional
optical response of the insulating media with metallic inclusions due to the so-called
surface plasmon (Mie, or geometric) resonances \cite{emt,Mie}, that have no
counterpart in homogeneous systems. These arise as a result of resonant behaviour of
local field in the granular composite system in presence of highly polarizable
inclusions and are governed to a considerable extent by the shape of the grains. The
frequency of geometric resonance is then easily obtained as the one at which the
polarizability of small particle diverges. For the case of spherical metallic
particles embedded in the insulating matrix with dielectric permittivity $\eps_d$ this
leads to the equation:
\begin{equation}
 \eps(\omega)_{part.}\,+\,2\,\eps_d\,=\,0,
\end{equation}
whence the resonance frequency is
\begin{equation}
  \omega_r\,=\,\frac{\omega_p}{\sqrt{1\,+\,2\,\varepsilon_d}},
\end{equation}
if the Drude's expression  with the plasma frequency $\omega_p$ is assumed for the
embeddings and $\eps_d$ is constant.

%-------------------------  FIG. 1. -----------------------
\begin{figure}
\includegraphics[width=\linewidth]{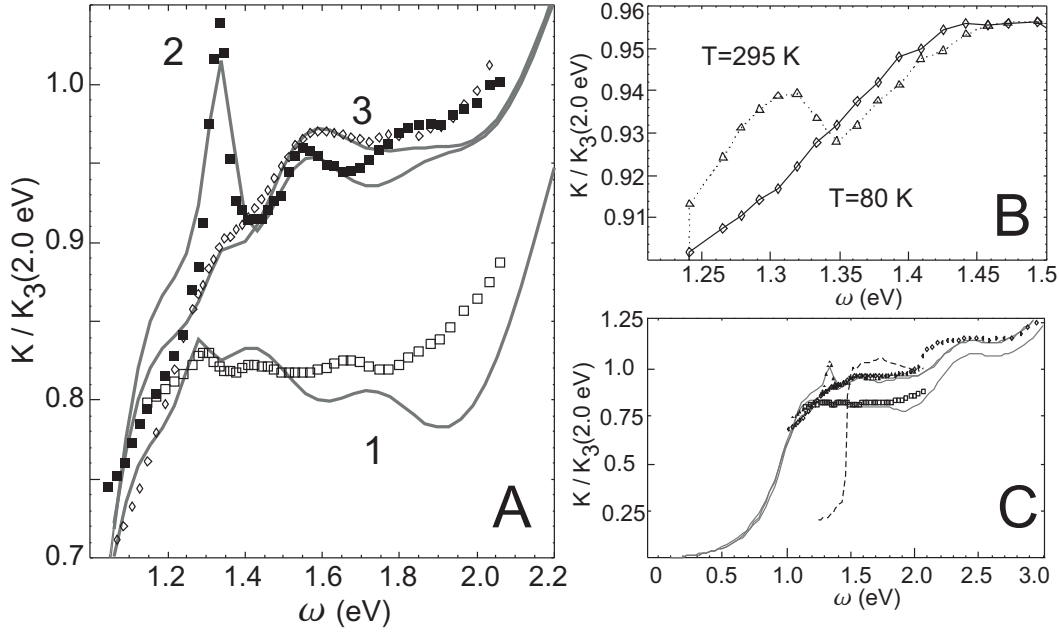}
\caption{A. Normalized room temperature absorption spectra of three CuO samples. Solid
lines are the effective medium theory fits (details are given in the text). B. A
fragment of the spectra of the sample 2, measured at two temperatures. C. The overall
evolution of the spectra in the whole range considered. Dashed line marks the
fundamental absorption band edge in pure CuO single crystal \cite{CuEl}.} \label{fig1}
\end{figure}
%-------------------------  FIG. 1. -----------------------

Turning to the case of nanocrystalline CuO samples, an additional complication of the
theoretical treatment imposed by the low symmetry of the monoclinic CuO unit cell is
to be faced. However, in the present study we may benefit by the fine-grained
structure of the samples, where all anisotropy is averaged out. Thus, in what follows,
we shall deal with the scalar effective permittivity $\eps_{eff}(\w)$. To calculate
this quantity in frames of effective medium theory, we made use of the familiar
Maxwell-Garnett-type approximation, widely applied to the description of heterogeneous
systems with the so called ''cermet topology'' (see, e.g. Ref. \cite{lamb}), where the
inclusions are completely surrounded by the host medium. It yields the following
expression for $\eps_{eff}$:
\begin{equation}\label{MG}
\eps_{eff}\,=\,\frac{(1 - p)\,\eps_d + \sum_i \Delta_i \eps_i}{(1 - p) + \sum_i
\Delta_i},
\end{equation}
where
\begin{equation}
\Delta_i\,=\,\left\langle \frac{\eps_i}{\eps_d + L^{\langle i \rangle}(\eps_i - \eps_d)} \right\rangle.
\end{equation}
Here $\eps_d$ denotes the dielectric permittivity of the bare
dielectric CuO medium, $\langle \cdots \rangle$ stands for the
averaging over the inhomogeneity, including the distribution of
the volume fractions $p_i$, orientations and other parameters,
pertaining to different species of embeddings with dielectric
permittivities $\eps_i$ and the principal values of the
depolarization tensors denoted by $L$'s, overall concentration of
inclusions being $p$. Thus, for a discrete set of species of
randomly oriented embeddings,
\begin{equation}\label{delta}
\Delta_i\, = \,\frac{p_i}{3} \sum \limits_{j=1}^{3} \frac{\eps_i}{\eps_d + L_j^{\langle i \rangle}(\eps_i - \eps_d)}, \quad \sum_i p_i = p,
\end{equation}
where $j$ runs over three depolarization factors of the grains of $i$-th species.
Given the effective dielectric permittivity, the absorption coefficient is defined by
the standard expression:
\begin{equation}
 K(\w)\, = \,2 \frac{w}{c} \mbox{Im} \sqrt{\eps_{eff}(\w)},
\end{equation}
It is useful to observe, that the working form of the above expression, where $K$ is
measured in cm$^{-1}$ and $\w$ in eV, is obtained by substitution $2/c \, \rightarrow
\,$ 101355.

The calculation of $\eps_{eff}$ with Eq.(\ref{MG}) requires the dielectric
permittivities of the components to be specified. In practice, we are forced to
make a number of strong simplifications and formulate a minimal model that
yields the explanation of main experimental findings. A simple reasonable model
implies the nanoceramic CuO to be  a system of identical metallic droplets with
volume fraction $p$ dispersed in an effective insulating matrix. We simulated
the dielectric CuO medium by the superposition of two Lorentz contributions:
\begin{equation}\label{lor}
 \eps_d\, = \,\eps_{\infty} + \frac{1}{\pi}\,\sum \limits_{k=1, 2} \frac{I_k}{\w_k^2 - \w^2 - \mbox{i}\,\Gamma_k\,\w},
\end{equation}
centered at $\hbar \omega_{1}=$ 2.35 eV ($\Gamma _{1}=$ 0.9 eV, $I_{1}=$ 14.256) and
$\hbar \omega_{2}=$ 4.5 eV ($\Gamma _{2}=$ 0.9 eV, $I_{1}=$ 469.8),
where $I_k$ and $\Gamma_k$ are the strength and the damping factor of $k$-th oscillator, respectively, and $\eps_{\infty}\, = \,$ 3.5. The permittivity of the metallic embeddings of $i$-th sort was described
by a simple Drude formula,
\begin{equation}
 \eps_i\, = \,1 - \frac{\Omega_{p,i}^2}{\w (w + \mbox{i} \gamma_i)}.
\end{equation}
This choice implies, that the characteristic transition energies of the novel
metallic-like phase lie well below those of the insulating one. An important
adjustable parameter, that  can substantially affect the absorption spectrum,
is the shape of inclusions, that enters the calculations through the
depolarization factors (Eq. \ref{delta}).  Assuming the shapes to be
ellipsoidal, we specified these with the ratios of two semi-axes  to the third
one, denoted by $\alpha$ and $\beta$, respectively.

Although the data in Fig. \ref{fig1} are presented in a dimensionless form, the
model numerical calculations are possible given the absolute value of the
absorption coefficient. However, as was pointed out in Sec. II, the artefact of
the experimental procedure, based on the transmittance measurements near the
fundamental band, results in a substantial underestimate of the magnitude of
the absorption. In order to provide the description of the optical spectra of
present cupric oxide nanoceramic, consistent with the experimental data
available for other cuprates, such as La$_2$CuO$_4$ \cite{uchida}, we
multiplied the measured absorption coefficient by an arbitrarily chosen factor
so as to set $K\,\sim\,10^5$ cm$^{-1}$ in the spectral range considered.

It is easy to see that choosing $p\approx 20 \%$, $\omega _p \approx
4\div 5$ eV, $\gamma \approx 0.5$ eV, $\alpha \approx 0.7$, $\beta
\approx 1.6$ we arrive at a reasonable description of the spectra of
all the samples excluding the "peak-dip-hump" feature revealed in
sample 2.  At first sight, its natural explanation might be related
to the exciton effects. However, this conventional approach
encounters troubles, because no similar feature of comparable
strength are observed in nominally pure CuO single crystals, and
above all, because of its quite unusual temperature dependence,
depicted in panel B of Fig. \ref{fig1}, where it is clearly seen, that peak
near 1.3 eV, being very sharp at the room temperature (295 K),
rapidly decays and finally disappears upon cooling to 80 K
that is  incompatible with the excitonic picture.

\begin{table*}
\caption{ Fit parameters of the metallic-like droplets for the
samples No. 1-3.}
\begin{center}
\begin{tabular}{c|ccccc|ccccc}
\hline
       No & $p_1$ & $\Omega_{p}^{\langle 1 \rangle}$  & $\gamma^{\langle 1 \rangle}$ &
       $\alpha_1$ & $\beta_1$  & $p_2$ & $\Omega_{p}^{\langle 2 \rangle}$  &
       $\gamma^{\langle 2 \rangle}$ & $\alpha_2$ & $\beta_2$\\ \hline

%        No      p1     Wp1    g1     a1      b1  |  p2      Wp2    g2     a2    b2
        1    &  0.186 & 4.35& 0.45  & 0.72 & 1.45  & 0.0265& 4.25 & 0.02 & 1.6 & 1.6 \\
        2    &  0.201 & 4.75& 0.5   & 0.73 & 1.6   & 0.0084& 4.75 & 0.01 & 1.9 & 1.9 \\
        3    &  0.20  & 4.75& 0.5   & 0.73 & 1.6   & 0.0042& 4.75 & 0.07 & 1.9 & 1.9  \\
%\hline
%       80  K &  0.123 & 5.2 & 0.33  & 1.25 & 1.625 & 0.022 & 4.65 & 0.13 & 2.1 & 2.1 \\
%       295 K &  0.122 & 5.2 & 0.323 & 1.25 & 1.625 & 0.024 & 4.65 & 0.13 & 2.1 & 2.1 \\
\hline
\end{tabular}
\end{center}
\end{table*}
At the same time, we believe, that the most plausible model of these
phenomena should take into account the correlation of the observed
spectral features and the effects of inhomogeneity in different
microgranular samples. The close examination of the spectra and
their comparison with the absorption spectrum of nominally pure
cuprates make it possible to conclude, that the inclusions present
in the samples are not monodisperse, but include several fractions
with distinctly different properties which simplified averaging can
provide only a "coarse-grained" description. In our approach, the
existence of rather narrow spectral feature, strongly marked in the
spectrum 2, points to a sizeable volume fraction of droplets with
"good" metallic properties, or small magnitude of damping parameter
$\gamma$. Among its most probable origins we consider e.g. the
precipitation of Cu nanograins under the influence of an intense
spherical shock waves, leading to the accumulation of numerous point
and linear defects. Indeed, the macroscopic concentration of Cu has
been previously found in cupric oxide after the fast particle
irradiation \cite{CuA} and plastic deformation \cite{9a}.
The measurements of the photoelectron Cu 2$p$ spectra (XPS) also
give evidence of a number of the copper ions Cu$^+$ with the reduced
valency, that amounts to a few percents, although this is at
variance to the XRD and XES data on the monophase structure of the
sample, reported above.
 Thus, to fit the fine resonant structure, clearly discernible in
the spectra 1, 2, 3, we introduced the simplest possible spread of the inhomogeneity
structure by considering two species of metallic-like droplets with volume fractions
$p_1 \gg p_2$ and damping parameters $\gamma _1 \gg \gamma _2$, or "bad" and "good"
metal, respectively. The fitted parameters for all the samples are listed in Table I.
Of course, the physical parameters of the particulate phases do not directly compare
with their ''bulk'' counterparts, e.g. due to the quantum size effect, that becomes
important for small particles in the size range below 100 nm \cite{emt}.

The comparison of the theoretical curves with experiments shows, that the resonant
structure of the absorption spectra, that evolve into the quite pronounced
"peak-dip-hump" feature in the spectrum 2, can be naturally assigned to the Mie
resonance.

We note the striking sensitivity of theoretical spectra to the
variation of the parameters of the inhomogeneity, and especially to
those of the phase with small damping constant. The slightest
changes in the relative volume fractions of two sorts of inclusions
result in a dramatic effects in the absorption spectra. This can
reasonably account for the difference in the spectra of the samples
2 and 3, because a subtle variation of the parameters from sample to
sample is practically unavoidable. Such a variation of the internal
texture of the inhomogeneity can be driven also with the
temperature. It is clearly observed in the panel B of the Fig.
\ref{fig1}, where a fragment of the absorption spectrum of the
sample 2 is depicted for two temperatures. Some discrepancy of the
present data with those of the main panel is due to the differences
in the experimental techniques, employed in both cases. It can be
clearly seen, that the pronounced peak at 1.34 eV in the
room-temperature spectrum completely disappears under cooling.
Interestingly, the spectra of the sample 2 measured at T=80 K look
quite similar to the room temperature spectra of the sample 2 and 3.
The effective medium approach makes it possible to reproduce the
essential features of experimental spectra slightly varying the
model parameters. The effect of the temperature can be ascribed
mainly to the changes in the relative volume of the second
(''good'') component. We have found, that the reduction of the
relative volume of the highly polarizable phasewith the lowering the
temperature by less than 8 $\%$ is enough to simulate such a
singular temperature behaviour. To get insight into the underlying
mechanism of this volume fraction effect, we note the onset of 3D
antiferromagnetic order in CuO, that takes place near $T\sim 230$ K.
In general, the onset of antiferromagnetism at low temperatures is
expected to entail the suppression of the metallic-like phase. The
way this occurs may be either the direct reduction of the volume
fraction and/or the changes in the properties of the droplet.
Moreover, the volume effect can be accompanied also by the
deformation of conducting inclusions, that results in some shift and
possible splitting of the geometric resonances, thus contributing to
the broadening of the feature, observed in the 295 K spectrum. We
regard its unusual temperature dependence as an important signature
to its ''geometric'' nature.

Within the conventional effective medium approach, the inclusions
are assumed to be much smaller, than the wavelength, and behave as
the point electric dipoles. No other Mie resonances besides the
first one (electric dipole) can be obtained within the model. To
take into account the finite size of the inclusions, higher order
terms of the Mie series are to be involved in the calculation of
$\eps_{eff}$. In a dilute limit, this step beyond the conventional
effective medium scheme can be implemented using the expression:
\begin{equation}\label{MGmie}
 \eps_{eff}\, = \, \eps_d \left(1 - \frac{3\, \mbox{i}\, p}{(\w/c)^3} S(0) \right), \;
 S(0)\, = \, \frac{1}{2} \sum \limits_n (a_n + b_n)
\end{equation}
where $S(0)$ is the forward scattering cross-section of spherical particle and the
Mie's coefficients $a_n,\,b_n$ are the functions of $\eps_d,\, \eps_i$ and the
particle diameter to the wavelength ratio, $x$. For $x \gtrsim 1$, the absorption
spectra acquire a fine ripple structure, related to the multipole excitations of the
spherical particle. We made use of the Eq.(\ref{MGmie}) to check on, whether the three
distinct observed feature of the absorption coefficient (Fig.\ref{fig1}) could be
related to the high-order Mie oscillations. It was found, that the observed features
could in fact be explained not only as due to three different dipole resonances of an
ellipsoidal particle, but as a result of the consecutive multipole resonances of a
spherical particle. However, in order the latter conclusion to be true, the improbably
large particles radii ($\sim$ 1550 nm) are required. A possible way to get round this
difficulty is to suggest the aggregation of the nanoscale inclusions. Recent numerical
extension of the Mie theory to the clusters of spherical inclusions \cite{ClustMie}
demonstrate the spectacular oscillations of the extinction coefficient of such
aggregates.

\section{Conclusion}
In conclusion, the optical absorption of nanocrystalline CuO samples
have been studied in spectral range $1\div 3$ eV. We found the
pronounced  spectral weight red-shift as compared with the spectra
of single crystalline CuO samples. In addition, the samples manifest
the fine spectral structure, such as a remarkable
temperature-dependent "peak-dip-hump" feature near $1.3\div 1.6$ eV.
The minimal model suggested to explain both effects implies the
nanoceramic CuO to be a system of two species of metallic-like
droplets with volume fractions $p_1 \gg p_2$ and damping parameters
$\gamma _1 \gg \gamma _2$, respectively, dispersed in an effective
insulating matrix. In other words, both effects are assigned to the
surface plasmon (Mie) resonances due to a small volume fraction of
metallic-like nanoscale droplets with Drude optical response
embedded in the bare insulating medium. Theoretical analysis within
the effective medium framework is shown to provide a good
semi-quantitative agreement with experiment. We underline the
striking sensitivity of the optical response to the variation of the
metallic volume fraction and other effective medium  parameters,
especially to those of the "good" metal phase. The slightest
unavoidable changes in the relative volume fractions of two sorts of
inclusions from sample to sample and with the temperature  result in
dramatic effects in the absorption spectra. Finally, we would like
to emphasize a clear similarity of the optical response produced by
the extrinsic nanoscale inhomogeneity created in different way (the
shock wave as in present study, the different irradiation) and
intrinsic nanoscale inhomogeneity created by a chemical doping in
parent insulating cuprates. This supports the idea of phase
instability of these SC oxides with regard to the nucleation of
metallic-like phase, or phase separation.

\ack The work was supported by INTAS 01-0654, CRDF No. REC-005,
Federal Program (Contract No. 40.012.1.1. 1153-14/02), grants RFBR
No. 01-02-96403, No. 02-02-96404, RFMC No. E00-3.4-280 and UR
01.01.042.

\end{document}